\documentclass{article}
\usepackage{spconf,amsmath,graphicx}

\usepackage{xcolor,soul,framed} 
\usepackage{array}
\usepackage{multirow}

\colorlet{shadecolor}{yellow}

\newcolumntype{P}[1]{>{\centering\arraybackslash}p{#1}}

\title{Fast Inter-Prediction based on Decision Trees for AV1 encoding}
\name{Jieon Kim*, Saverio Blasi$^\circ$, Andre Seixas Dias$^\circ$, Marta Mrak$^\circ$ and Ebroul Izquierdo* } 
\address{* Queen Mary University of London (UK) \\ $^\circ$BBC R\&D, London (UK)}
\begin{document}
%
\maketitle
\begin{abstract}
The AOMedia Video 1 (AV1) standard can achieve considerable compression efficiency thanks to the usage of many advanced tools and improvements, such as advanced inter-prediction modes. However, these come at the cost of high computational complexity of encoder, which may limit the benefits of the standard in practical applications. This paper shows that not all sequences benefit from using all such modes, which indicates that a number of encoder optimisations can be introduced to speed up AV1 encoding. A method based on decision trees is proposed to selectively decide whether to test all inter modes. Appropriate features are extracted and used to perform the decision for each block. Experimental results show that the proposed method can reduce the encoding time on average by 43.4\% with limited impact on the coding efficiency. 
\end{abstract}
\section{Introduction}
\label{sec:intro}

The Alliance for Open Media (AOMedia) \cite{aom} recently finalised the development of the AOMedia Video 1 (AV1) specification. AOMedia was founded in 2015 as a consortium of over 30 partners from the semiconductor industry, video on demand providers and web browser developers, with the specific objective of creating open, royalty-free multimedia delivery solutions. AV1 is the first outcome of such initiative, and it was built using the VP9 standard specification developed by Google \cite{VP9} as a base. Similarly to its predecessor, AV1 follows the typical hybrid block-based approach commonly used in many video coding standards. 

Thanks to many new optimisations, algorithms and techniques, AV1 has significantly better performance in terms of higher quality with considerable bit rate saving compared to VP9 \cite{t_laude}. This is mostly due to the fact that AV1 adopts a number of new technical contributions, such as enhanced directional intra-prediction, extended reference frames, dynamic spatial and temporal motion vector referencing, overlapped block motion compensation, extended transform kernels, and many others \cite{ov_av1}. While these large number of tools and encoder options contribute to the compression efficiency of the standard, encoder implementations are required to select the best configuration for each portion of the sequence being encoded. This comes at the cost of considerable additional computational complexity, which may limit the benefits of using the standard in practical applications \cite{ibcAndre}. Therefore, algorithms to reduce the encoder run time with limited effects on the standard coding efficiency would be highly beneficial. 

In this paper, a method to reduce the complexity of an AV1 encoder based on early termination of inter-prediction is presented. The method is based on machine learning techniques in order to reduce the number of options to test at the encoder side. The rest of the paper is organised as follow. Section \ref{sec:state_of_the_art} briefly presents state-of-the-art encoder speed-up techniques which make use of machine learning. Section \ref{sec:motivation} provides an overview of AV1 inter-prediction, as well as the motivation of the proposed method. Then, Section \ref{sec:proposed} presents the proposed binary tree based inter mode decision algorithm. Experimental results and analysis are presented in Section \ref{sec:results}. Finally, conclusions are drawn in Section \ref{sec:conclusions}.

\section{State of the art}
\label{sec:state_of_the_art}

Due to the fact that AV1 was recently finalised, limited work is found in the literature related to reducing the complexity of AV1. A paper was presented focusing on predicting the optimal block size of AV1 encoding based on Bayesian inference \cite{guo_av1}. In addition, some work was proposed to speed up AV1 encoding in multi-rate configurations, exploiting information obtained in one representation to speed up the encoding of the other representations \cite{guo_av1_multirate}. 

AV1 follows a similar architecture to standards developed by the ITU-T VCEG and/or ISO/IEC MPEG, and as such, it is relevant to briefly present some of the speed-up tools based on machine learning that were proposed in this context. Shen et al. \cite{l_shen} proposed an early termination algorithm for transform block size determination in the High Efficiency Video Coding (HEVC) standard, where the Bayesian decision theory was applied to map the variance of the residual coefficients to the block size. In \cite{g_correa} the decision trees generated by data mining tools were utilised to predict the size of HEVC blocks. Furthermore, a method was proposed to limit the number of block sizes to test in HEVC based on exploiting the size of neighbouring blocks \cite{zupancic}. Similarly, a method to select the optimal motion vector precision was proposed \cite{blasi}, based on local features, such as the behaviour of the residual error samples, and global features, such as the amount of edges in the pictures. These methods were proposed in the context of different codecs and may not be applicable to apply directly to the AV1 coding structure. 

\section{Motivation}
\label{sec:motivation}

In AV1, an inter-predicted block can be encoded with either Single Reference Frame Prediction Mode (SRFPM), in which case one single reference frame with a corresponding motion vector is used for the prediction, or with Compound Reference Frame Prediction Mode (CRFPM), where two reference frames (with two corresponding motion vectors) are used. Up to seven reference frames can be used by either mode, referred to as LAST\_FRAME, LAST2\_FRAME, LAST3\_FRAME, GOLDEN\_FRAME, BWDREF\_FRAME, ALTREF\_FRAME, and ALTREF2\_FRAME.
More details on this selection can be found in the literature \cite{ov_av1}.
\begin{figure}[!bt]
	\centering 
	\includegraphics[scale=0.25]{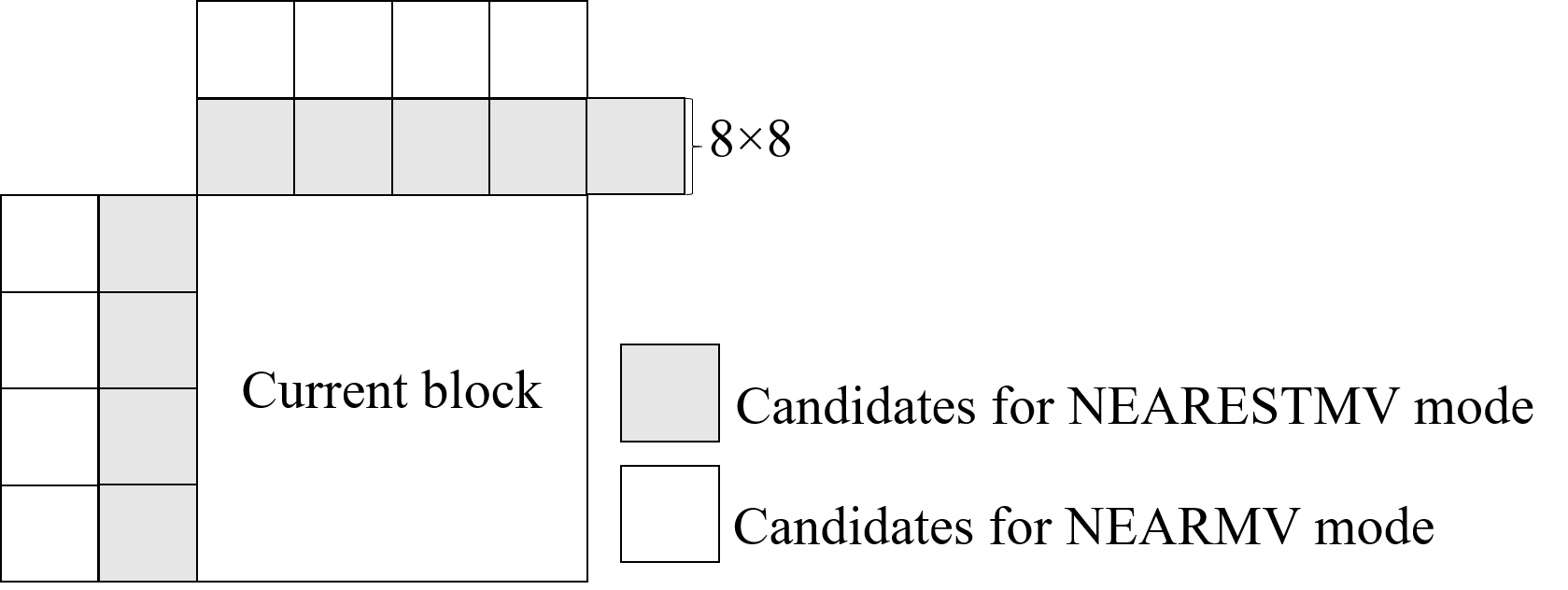}
	\caption{NEARESTMV and NEARMV modes}\label{fig:nearestmv}
	\vspace*{-6mm} 
\end{figure}

Four motion vector candidates are used in SRFPM, which are NEARESTMV, NEARMV, NEWMV and GLOBALMV. NEARESTMV and NEARMV modes employ previously coded motion vectors extracted from spatial neighbours, as shown in Fig.~\ref{fig:nearestmv}. In addition, NEWMV mode performs block based motion estimation to generate a new motion vector for the current block, while GLOBALMV mode performs frame based motion estimation to generate a single motion vector candidate for the whole frame. Eight candidates are used in CRFPM, obtained by combining some of the candidates in SRFPM to perform bi-directional inter-prediction.

In conventional AV1 encoder implementations, the encoder can select among all these different options, the best option for the current block. Typical implementations base these decisions on rate-distortion optimisation techniques, in which the options are compared based on a cost that takes into account the number of bits needed to encode the block, and the corresponding distortion. Clearly, exhaustively searching among all these options can lead to significant encoder complexity. 

\begin{figure}[!hbt]
	\centering 
	\includegraphics[width=8cm,height=4.5cm]{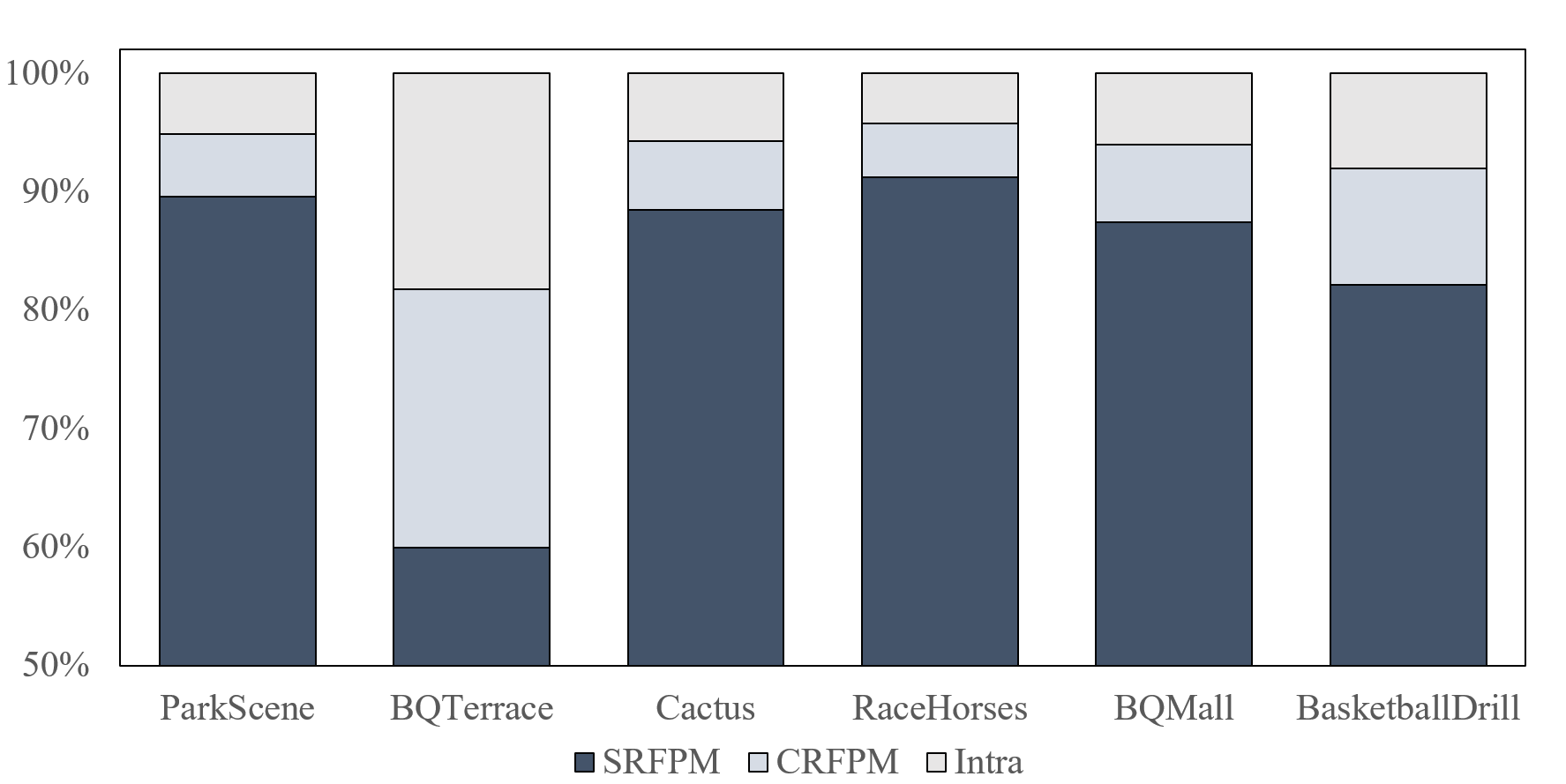}
	\caption{Percentage of the best prediction modes in AV1}\label{fig:bestpumodes_av1}
		\vspace*{-1mm} 
\end{figure}

On the other hand, it is likely that different types of content may benefit from different coding modes. If the encoder could identify which options to use without performing a brute-force search, considerable time savings could be obtained with limited impact on the compression efficiency. Hence, a statistical analysis was performed to analyse which modes are mostly used in specific sequences, as shown in Fig.~\ref{fig:bestpumodes_av1}. The figure presents the average occurrence probability  of  SRFPM,  CRFPM  and  intra-prediction  in  blocks extracted  from  inter  frames in AV1 software (Mar, 2018 version), where four different QPs (32,  43,  55 and  63) considered. As can be seen, most inter-predicted blocks are encoded mainly with SRFPM. For example, around 91\% and 60\% of inter-blocks are encoded with SRFPM in sequences \textit{RaceHorses} and \textit{BQTerrace} respectively. 

Furthermore, the complexity and coding efficiency of using CRFPM were analysed. The AV1 encoder was modified to prevent testing and selection of CRFPM. This modified encoder was compared with an anchor, namely a conventional AV1 encoder that can select CRFPM. The compression performance of the modified encoder with respect to the anchor was measured in terms of the well-known BD-BR metric,  a measure of the difference in rate required to encode at the same objective quality with respect to the anchor at different quality points, in percentage \cite{bdbr}. Complexity was measured in terms of the difference in encoding time, calculated as:
\begin{equation}
  TS[\%] = (1 -  \frac{T(C\textsubscript{modified})}{T(C\textsubscript{anchor})})\cdot 100,
\end{equation}
where $T(C\textsubscript{anchor})$ and $T(C\textsubscript{modified})$ are the total encoding times required by the anchor and the modified encoder, respectively. Results of this test are presented in Table ~\ref{table:ccr}. 
 As can be seen, avoiding testing CRFPM modes reduce complexity by average 61.4\% in terms of encoding time. On the other hand, forcing the encoder to simply remove CRFPM modes can have a  detrimental effect on encoding performance in some cases, with up to $12.7\%$ efficiency losses in the case of the \textit{BQTerrace} sequence. In this paper, a method to selectively predict blocks in which CRFPM would be needed is presented using data mining. The proposed method can be considered as a binary-class classification task which is applied to each block to decide whether it should be predicted using SRFPM (Class 0) or using either CRFPM or  SRFPM (Class 1).

\begin{table}[hbt]
\centering
\caption{Complexity and efficiency of CRFPM modes}
\resizebox{.9\linewidth}{!}{
\begin{tabular}{l|l|c|c}
\hline
\hline
\multicolumn{1}{c|}{\textbf{Sequence}} & \multicolumn{1}{c|}{\textbf{Resolution}} & \textbf{TS [\%]} & \textbf{BD-BR [\%]} \\ \hline
ParkScene & \multirow{3}{*}{1920 $\times$ 1080} & 62.8 & 0.55 \\ \cline{1-1} \cline{3-4} 
BQTerrace &  & 66.7 & 12.73 \\ \cline{1-1} \cline{3-4} 
Cactus &  & 50.7 & 1.62 \\ \hline\hline
RaceHorses & \multirow{3}{*}{832$\times$480} & 60.4 & 0.73 \\ \cline{1-1} \cline{3-4} 
BQMall &  & 64.6 & 2.48 \\ \cline{1-1} \cline{3-4} 
BasketballDrill &  & 63.0 & 1.94 \\ \hline

\multicolumn{2}{c|}{\textbf{Average}} & \multicolumn{1}{c|}{\textbf{61.4}} & \multicolumn{1}{c}{\textbf{3.34}} \\ \hline

\hline
\end{tabular}}
\label{table:ccr}
\end{table}

\section{Inter-coding early termination based on decision trees}
\label{sec:proposed}

Decision trees are simple yet an effective tool to learn the relationship between a set of features and the ground truth. A decision tree is a hierarchical structure consisting of a group of decision nodes and terminal leaves, where each node corresponds to a specific test on a single feature, and the terminal leaves provide a classification for the ground truth. The work in this paper made use of a well-known open-source implementation \cite{cusplit} for training the decision trees. 

Different from more complex machine learning solutions, the application of trained decision trees has the advantage of being very simple to implement, leading to little additional complexity. This is crucial from the problem at hand of reducing the complexity of an AV1 encoder implementation. Using more complex solutions such as methods based on support vector machines or deep learning may lead to better classification results, but this comes at the cost of high complexity of applying the method itself during encoding. Given the decision needs to be taken for each inter-predicted block during encoding, such complexity would have a detrimental impact on encoding time, compromising the effectiveness of the algorithm. 

As with all machine learning techniques, the selection of sequences used for the training is crucial to ensure that the method can generalise well. To this aim, the first 20 frames of a set of well-known video sequences used in the development of MPEG standards \cite{ctc} was used, as shown Table ~\ref{table:training_sequences}. Motion activity, texture and resolution are different for the selected test sequences. 

\begin{table}[!bt]
\centering
\caption{Sequences for training data set}
\resizebox{.8\linewidth}{!}{
\begin{tabular}{l c c c}
\hline
\hline
\multicolumn{1}{c}{\textbf{Sequence}} & \textbf{Frame Rate} & \textbf{Bit Depth} & \textbf{Resolution} \\ \hline
ParkScene & 24 & 8 & 1920 $\times$ 1080 \\ \hline
BQTerrace & 60 & 8 & 1920 $\times$ 1080 \\ \hline
Kimono1 & 24 & 8 & 1920 $\times$ 1080 \\ \hline
Cactus & 50 & 8 & 1920 $\times$ 1080 \\ \hline
RaceHorses & 30 & 8 & 832 $\times$ 480 \\ \hline
BQMall & 60 & 8 & 832 $\times$ 480 \\ \hline
PartyScene & 50 & 8 & 832 $\times$ 480 \\ \hline
BasketballDrill & 50 & 8 & 832 $\times$ 480 \\ \hline
\hline
\end{tabular}
}
\label{table:training_sequences}
\end{table}
Also important to the accuracy of a machine learning algorithm is the selection of features to use for the classification. Ideally the features should be highly correlated to the ‘ground truth’, which in the case under examination is whether a block is encoded using SRFPM or CRFPM, but not very correlated with each other. This minimises the inclusion of unnecessary data and inter-feature correlations. To this end, many features were extracted from each block in the training sequences. Each feature was then classified in terms of its Gini impurity with respect to the ground truth. The Gini impurity is a measurement of the likelihood of making an incorrect classification of a new instance of a random variable, and it is calculated as follows:
\begin{equation}
 G = \sum_{i=0}^{N-1}[{{{P(i)}\cdot{(1-P(i))}}]},
\end{equation}
where $P(i)$ is the probability of block \textit{i} being encoded with SRFPM, and \textit{i} indicates blocks in the data training set.

Following this process, four features were selected, which aims at 80\% accuracy of the prediction in Class 0. Table ~\ref{table:features} shows the selected features. Two features are taken from each of the adjacent encoded blocks, on the left and on the top of the current block. These features are denoted as 'f\textsubscript{1}' and 'f\textsubscript{2}' for the left block, and 'f\textsubscript{3}' and 'f\textsubscript{4}' for the top block. They have been selected because there is a high correlation in coded information between the current block and its neighbours. 

\begin{table}[!hbt]
\centering
\caption{Selected features and descriptions}
\resizebox{.8\linewidth}{!}{
\begin{tabular}{p{2cm} p{5.5cm}}
\hline
\hline
\textbf{Feature} & \textbf{Description} \\  \hline
f\textsubscript{1} & second reference frame of left block \\ \hline
f\textsubscript{2} & prediction mode of left block \\ \hline
f\textsubscript{3} & second reference frame of upper block \\ \hline
f\textsubscript{4} & prediction mode of upper block \\ \hline
\end{tabular}}
\label{table:features}
\end{table}

Given that the sequences in the training set are of different length and different resolutions, using all blocks in each sequence would lead to an unbalanced number of training samples from each sequence. Sequences at high resolutions may therefore have a higher impact on the training, which is not ideal to ensure the training can generalise well. Therefore, a fixed number of training samples is used per sequence. Moreover, in order to balance the training samples between the two classes of the ground truth, the number of training samples \textit{M} for each sequence is calculated as 
\begin{align*}
M=\left\{\begin{array}{@{}cc}
p\textsubscript{0}\cdot N,&p\textsubscript{0} < 0.5\\
(1-p\textsubscript{0})\cdot N,&p\textsubscript{0} \geq 0.5,
\end{array}
\right.
\end{align*}
where $p\textsubscript{0}$ represents the accuracy of SRFPM prediction, and $N$ corresponds to the number of training samples in each class. Hence, the data set is balanced with 50\% of blocks classified as being predicted using SRFPM, and 50\% using CRFPM. 

\begin{figure}[hbt]
	\centering 
	\includegraphics[scale=0.35]{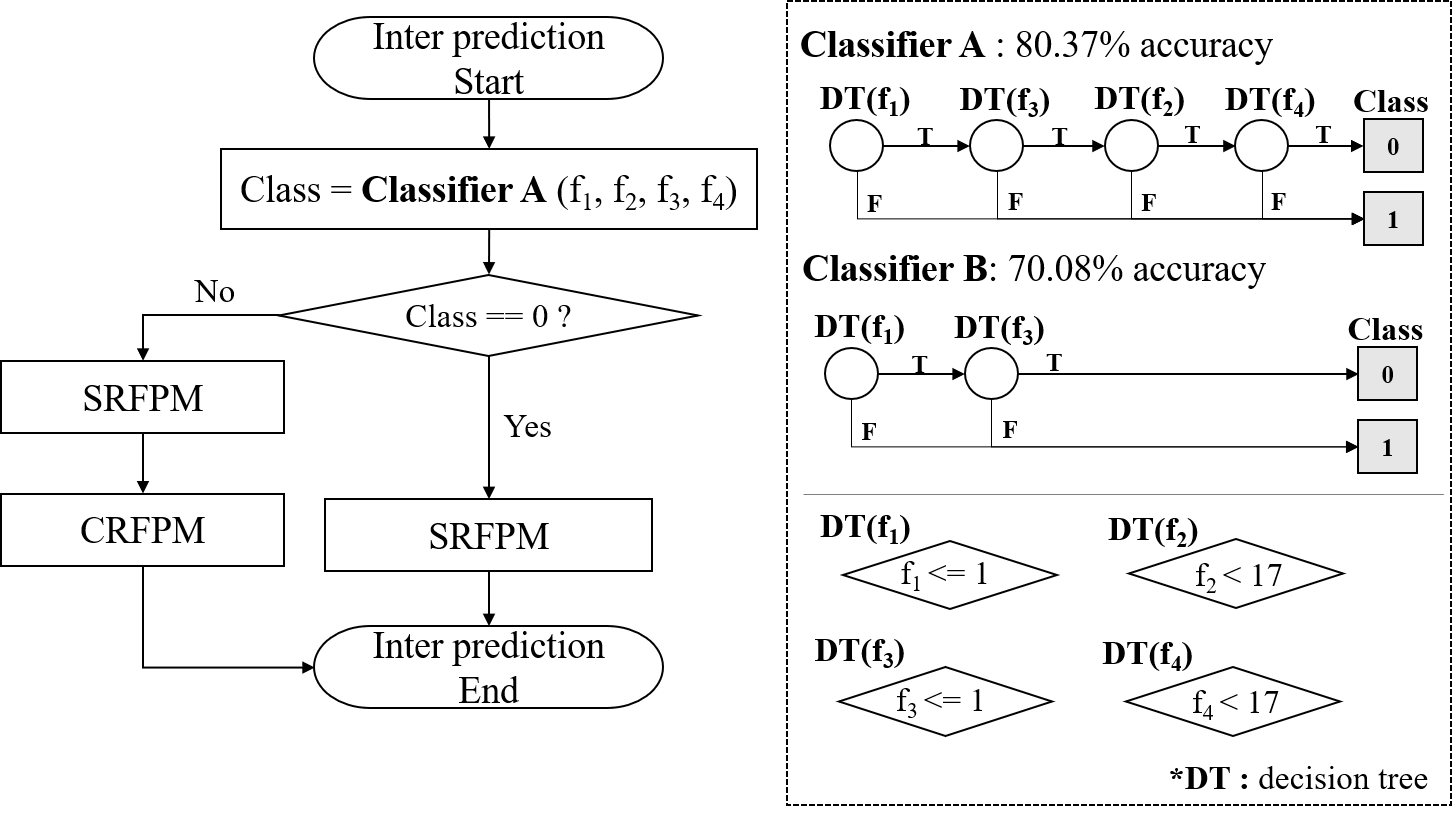}
	\caption{Flowchart of the proposed method}\label{fig:flowchart}
\end{figure}

The flowchart of the proposed method is illustrated in Fig.~\ref{fig:flowchart}. A binary classifier (Classifier A) is firstly used to make a decision per block. If the block is classified as Class 0, which corresponds to the block being predicted using SRFPM,  then CRFPM modes are not tested for the current block. Conversely, if the block is classified in Class 1, both SRFPM and CRFPM modes are tested as in conventional AV1 encoders.

\section{Experimental Results}
\label{sec:results}

The method was tested to evaluate its performance using test sequences that are not part of the training used to develop the method. In order to validate the performance of the proposed method, an encoder was developed using the reference AV1 software (Sep 5, 2018 version) encoder as a basis \cite{av1}. The unmodified reference software encoder was also used as anchor for measuring performance. Seven different video sequences \cite{testset} were used to evaluate the performance. Each sequence was encoded at four different quality points (obtained using  --cq-level=32, 43, 55, 63), to validate the method under different conditions. The performance of the proposed method was measured in terms of BD-BR and encoder time savings. Table ~\ref{table:summaryresults} shows that the proposed method reduces encoding time on average by 43.4\%, with a maximum time saving of 53.0\% and a minimum of 33.7\%. As can be seen, the method achieves considerably better coding efficiency than the modified encoder which skips testing of CRFPM modes altogether. 

\begin{table}[!bt]
\centering
\caption{Performance of the proposed approach and C\textsubscript{modified}}
\resizebox{\linewidth}{!}{
\begin{tabular}{l|c|c|c|c}
\hline\hline
\multirow{2}{*}{\textbf{Sequence}} & \multicolumn{2}{c|}{\textbf{Proposed method}} & \multicolumn{2}{c}{\textbf{C\textsubscript{modified}}} \\ \cline{2-5}  & \textbf{BD-BR[\%]} & \textbf{TS[\%]} & \textbf{BD-BR[\%]} & \textbf{TS[\%]} \\ \hline
blue\_sky\_360p\_120f & 0.62 & 45.1 & 1.04 & 62.4 \\ \hline
old\_town\_cross\_1080p50\_60f & 0.28 & 53.0 & 1.54 & 59.4 \\ \hline
pedestrian\_area\_1080p25\_60f & 0.35 & 38.5 & 0.54 & 58.5 \\ \hline
speed\_bag\_640x360\_120f & 1.00 & 33.7 & 4.71 & 59.3 \\ \hline
stockholm\_640x360\_120f & 0.14 & 50.5 & 0.13 & 66.5 \\ \hline
tacomanarrows360p\_120f & 0.62 & 46.8 & 0.72 & 61.0 \\ \hline
thaloundeskmtg360p\_120f & 2.37 & 35.7 & 10.97 & 60.1 \\ \hline
\textbf{Average} & \textbf{0.77} & \textbf{43.4} & \textbf{2.81} & \textbf{61.0} \\ \hline\hline
\end{tabular}}
\label{table:summaryresults}
\end{table}

The AV1 reference software allows encoding to be performed using a variety of so called "Speed presets", namely encoder configurations which limit certain options and tools in order to reduce the complexity. When doing so, the coding efficiency decreases due to the fact the codec is limited in the number of options it can select. On the other hand, the proposed method selects whether to test or not the CRFPM modes on a block-by-block basis based on features of the block, and as such it has a limited impact on coding efficiency. Moreover, the tool can be used on top of existing AV1 Speed presets and still provide additional speed-ups, showing that the method does not overlap with existing complexity reduction schemes. In order to validate these claims, the method was compared with AV1 Speed preset $2$, and it was also tested on top of an AV1 encoder using Speed preset $2$. 

Results of these tests are presented in Table ~\ref{table:summaryresults2}. In all cases, the unmodified AV1 encoder was used as anchor. As can be seen, using AV1 Speed preset $2$ on its own usually provides lower complexity reductions for higher efficiency losses than using the proposed method (as in Table ~\ref{table:summaryresults}), showing that the Speed preset is less capable of adapting to content-dependent features. Moreover, using the proposed method on top of Speed preset $2$ can still provide considerable complexity reductions, showing that the method is almost orthogonal to the Speed preset. Average $58\%$ and up to $64\%$ complexity reduction can be obtained under these conditions.

\begin{table}[!bt]
\vspace*{-5mm} 
\centering
\caption{Performance of Speed preset $2$, and the proposed approach used in combination with Speed preset $2$}
\resizebox{\linewidth}{!}{
\begin{tabular}{l|c|c|c|c}
\hline\hline
\multirow{2}{*}{\textbf{Sequence}} & \multicolumn{2}{P{2.3cm}|}{\textbf{Speed preset $2$}} & \multicolumn{2}{p{2.7cm}}{\textbf{Proposed method + Speed preset $2$}} \\ \cline{2-5}  & \textbf{BD-BR [\%]} & \textbf{TS [\%]} & \textbf{BD-BR [\%]} & \textbf{TS [\%]} \\ \hline
blue\_sky\_360p\_120f & 0.45\ & 32.5\ & 1.19\ & 54.8\ \\ \hline
old\_town\_cross\_1080p50\_60f & 0.66\ & 33.5\ & 1.11\ & 60.9\ \\ \hline
pedestrian\_area\_1080p25\_60f & 0.58\ & 40.5\ & 0.73\ & 55.0\ \\ \hline
speed\_bag\_640x360\_120f & 1.00\ & 41.2\ & 1.82\ & 55.1\ \\ \hline
stockholm\_640x360\_120f & -0.56\ & 33.2\ & 0.05\ & 60.3\ \\ \hline
tacomanarrows360p\_120f & 1.20\ & 48.6\ & 1.34\ & 64.2\ \\ \hline
thaloundeskmtg360p\_120f & 2.94\ & 34.6\ & 4.88\ & 53.6\ \\ \hline
\textbf{Average} & \textbf{0.90}\ & \textbf{37.7}\ & \textbf{1.59}\ & \textbf{57.7} \\ \hline\hline
\end{tabular}}
\label{table:summaryresults2}
\end{table}

\section{Conclusions}
\label{sec:conclusions}

This paper presented a fast algorithm for AV1 inter prediction based on decision trees, based on the observation that not all sequences benefit from using CRFPM modes. A decision tree was trained based on $7$ features extracted while encoding each block. A decision is performed whether to skip testing of CRFPM modes, or whether to instead test all modes exhaustively. Experimental results show that the  encoding time can be reduced on average by 43.4\%, with negligible impact on coding efficiency. Future work could focus on using more features and using different classifier to further speed up AV1 encoder implementations.



\bibliographystyle{IEEEbib}
\bibliography{refs}

\begin{thebibliography}{10}

\bibitem{aom}
``Aom - alliance for open media,'' http://aomedia.org/.

\bibitem{VP9}
D.~Mukherjee, J.~Bankoski, A.~Grange, J.~Han, J.~Koleszar, P.~Wilkins, Y.~Xu,
  and R.S. Bultje,
\newblock ``The latest open-source video codec vp9 - an overview and
  preliminary results,''
\newblock in {\em Picture Coding Symposium (PCS)}, 2013.

\bibitem{t_laude}
T.~Laude, Y.G. Adhisantoso, J~.Voges, M.~Munderloh, and J.~Ostermann,
\newblock ``A comparison of {JEM} and {AV1} with {HEVC}: {Coding Tools, Coding
  Efficiency and Complexity},''
\newblock in {\em Picture Coding Symposium (PCS)}, June 2018.

\bibitem{ov_av1}
Y.~Chen et~al.,
\newblock ``{An Overview of Core Coding Tools in the AV1 Video Codec},''
\newblock in {\em Picture Coding Symposium (PCS)}, June 2018.

\bibitem{ibcAndre}
A.~S. Dias, S.~Blasi, F.~Rivera, E.~Izquierdo, and M.~Mrak,
\newblock ``{An overview of recent video coding developments in MPEG and
  AOMedia},''
\newblock in {\em International Broadcasting Convention (IBC)}, September 2018.

\bibitem{guo_av1}
B.~Guo, Y.~Han, and J.~Wen,
\newblock ``Fast block structure determination in av1-based multiple
  resolutions video encoding,''
\newblock in {\em 2018 IEEE International Conference on Multimedia and Expo
  (ICME)}, July 2018, pp. 1--6.

\bibitem{guo_av1_multirate}
B.~Guo, X.~Chen, J.~Gu, Y.~Han, and J.~Wen,
\newblock ``A bayesian approach to block structure inference in av1-based
  multi-rate video encoding,''
\newblock in {\em 2018 Data Compression Conference}, March 2018, pp. 383--392.

\bibitem{l_shen}
L.~Shen, Z.~Zhang, X.~Zhang, P.~An, and Z.~Liu,
\newblock ``Fast tu size decision algorithm for hevc encoders using bayesian
  theorem detection,''
\newblock in {\em Signal Process. Image Commun.}, March 2015, vol.~32, pp.
  121--128.

\bibitem{g_correa}
G.~Correa, P.~A. Assuncao, L.~V. Agostini, and L.~A. da~Silva~Cruz,
\newblock ``{Fast HEVC encoding decisions using data mining},''
\newblock in {\em IEEE Trans. Circuits Syst. Video Technol.}, Apr. 2015,
  vol.~25, pp. 660--673.

\bibitem{zupancic}
I.~Zupancic, S.~G. Blasi, E.~Peixoto, and E.~Izquierdo,
\newblock ``Inter-prediction optimizations for video coding using adaptive
  coding unit visiting order,''
\newblock {\em IEEE Transactions on Multimedia}, vol. 18, no. 9, pp.
  1677--1690, Sept 2016.

\bibitem{blasi}
S.~G. Blasi, I.~Zupancic, E.~Izquierdo, and E.~Peixoto,
\newblock ``Adaptive precision motion estimation for hevc coding,''
\newblock in {\em 2015 Picture Coding Symposium (PCS)}, May 2015, pp. 144--148.

\bibitem{bdbr}
G.~Bjontegaard,
\newblock ``{Calculation of average PSNR differences between RD-Curves},''
\newblock in {\em ITU-T SG16 Q.6 Document, VCEG-M33}, April 2001.

\bibitem{cusplit}
``Decision tree classification tool for video coding optimisation problems,''
  https://github.com/bbc/cu\_split.

\bibitem{ctc}
F.~Bossen,
\newblock ``Common test conditions and software reference configurations,''
\newblock in {\em JCTVC-L1100}, October 2012.

\bibitem{av1}
``Av1 source code in the alliance for open media git repository,''
  https://aomedia.googlesource.com/aom/.

\bibitem{testset}
``objective-1-fast test set in awcy,''
  https://people.xiph.org/\%7etdaede/sets/objective-1-fast/.

\end{thebibliography}

\end{document}